
\magnification\magstep1
\baselineskip=16pt

\def\Bpsi{{\bar \Psi}}

\def\a{\alpha}

\def\no{\noindent}
\def\en{\eqno}
\no
\centerline{\bf An Exactly Soluble Model of Directed Polymers}
\centerline{\bf with Multiple Phase Transitions}\par
\vskip 0.2in
\centerline{G. Forgacs$^1$ and K. Ziegler$^2$}
\vskip 0.3in
\centerline{$^1$Department of Physics, Clarkson University}
\centerline{Potsdam, N.Y. 13699-5820, USA}
\centerline{and}
\centerline{$^2$Institut f\"ur Theorie der Kondensierten Materie,
Universit\"at,}
\centerline{ Physikhochhaus, D-76128 Karlsruhe, Germany}
\vskip 0.6in
\noindent Abstract:\par
\no
Polymer chains with hard-core interaction on a two-dimensional lattice are
modeled by directed random walks. Two models, one with intersecting walks (IW)
and another with
non-intersecting walks (NIW) are presented, solved and compared. The exact
solution of the two models,
based on a representation using Grassmann variables, leads, surprisingly, to
the same analytic
expression for the polymer density and identical phase diagrams. There are
three different phases as a
function of hopping probability and single site monomer occupancy, with a
transition from the dense
polymer system to a polymer liquid (A) and a transition from the liquid to an
empty lattice (B).
Within the liquid phase there exists a self-dual line with peculiar properties.
The derivative of
polymer density with respect to the single site monomer occupancy diverges at
transitions A and B,
but is smooth across and along the self-dual line. The density-density
correlation function along the
direction $x$, perpendicular to the axis of directedness has a power law decay
1/$x^2$ in the
entire liquid phase, in both models. The difference between the two models
shows up
only in the behavior of the correlation function along the self-dual line: it
decays
exponentially in the IW model and as 1/$x^4$ in the NIW model.

\vfill \eject \no
A directed random walker makes steps along a given ($z$) axis only in one
(forward) direction. Random fluctuations are present in the transverse
directions$^1$. Such walks are the subject of great interest since they may
model
diverse phenomena like polymers under flow$^2$, tracer diffusion$^3$,
electrorheological
fluids$^4$, commensurate - incommensurate phase transitions$^5$, vortex lines
in high
$T_c$ superconductors$^{6,7}$, world lines of quantum particles$^8$, the
behavior of interfaces in $1+1$ dimensions$^9$, some aspects of
biomembranes$^{10}$, etc. They share a number of properties of dimers$^9$ and
vortex models$^{11}$.

Directedness makes analytic calculations considerably easier, and a number
of exact results exist both for a single walk and cases with many walks.

In the present work we concentrate on some generic properties of random walks.
We
introduce and solve exactly a model with non-intersecting walks. Using the same
formalism, we also obtain the solution of a similar model, which allows
for intersections. The comparison of the two models reveals some surprising
effects.  The
interesting feature of our systems is that along with second order phase
transitions, observed in previous investigations of similar models$^{12-14}$,
they
contain a self-dual line, with characteristics of yet another critical
phenomenon.

We first discuss the NIW model depicted in Fig.1. Chains are pinned
with one of their ends at $z=0$. In order to avoid crossing of walks in
the vertical direction only steps of length $2a$ are allowed, where $a$
is the lattice spacing. Each such chain can be parametrized with two
indices, $r$ and $j$. Here $r$ are lattice sites denoted by open circles in
Fig.1. The index $j$ takes values 0 or 1. $j=0$ if a site is an $r$-site,
and, as a consequence, can be reached from another $r$-site making two steps
along the
elementary vectors $e_1$ (site $O_{\rm 1}$ in Fig.1) or $e_3$ (site
$O_{\rm 2}$ in Fig.1). $j=1$ for a site which
can be reached from an $r$-site by making a single step in the direction of the
vector
$e_3$ (site $O_{\rm 3}$ in Fig.1) or from a $j=1$ site directly below (site
$O_{\rm 4}$ in Fig.1). Chains start at $z=0$, at $r$-sites (also needed to
avoid crossing). This is model A of ref. 14 which was used to
describe phase transitions in biomembranes. In the present article we
investigate correlations in this model (as well as in the IW model) to provide
a deeper understanding
of the statistics than what can be obtained from the thermodynamic properties
studied in
ref. 14.

We impose periodic boundary
conditions both in the $x$ and $z$ directions. We assume a hard-core potential
between walks: chains repel each other with infinite energy upon contact.
Therefore, the contribution of the configuration in Fig.1 to the partition
function is zero because of point $O_{\rm 4}$. We assign unit weight to a step
in the vertical direction (with length $2a$ along the vector $e_1$), and a
weight $w$ to a step along the diagonals of the elementary square. The
weight (fugacity) of an empty site is denoted by $\mu$. Introducing a pair of
Grassmann variables$^{15}$ $\{\Psi_{r,j},\Bpsi_{r,j}\}$ for each lattice site,
the configurations of our system can be generated from the partition function

$$Z=\int\exp(\sum_{r,r'}\sum_{j=0,1}\Psi_{r,j}G^{-1}_{r,r';j,j'}\Bpsi_
{r',j'}){\cal D}\Psi{\cal D}\Bpsi={\rm det} G^{-1}.\en (1)$$

Here ${\cal D}\Psi{\cal D}\Bpsi$ denotes the Berezin integration$^{15}$ over
all Grassmannians, and, the lattice Green's function is given by its Fourier
components

$${\tilde G}^{-1}_k=\pmatrix{
e^{ik_1} -\mu&w e^{ik_1}(1+e^{ik_2})\cr
w(1+e^{-ik_2})& e^{ik_1}-\mu\cr
}.\en (2)$$

\no The $2\times2$-matrix structure reflects the dependence on the index
$j=0,1$; $k_1$ and $k_2$ are the components of  the two-dimensional momentum
vector
along the $x$- and $z$-directions, respectively. On a lattice with $N$ sites
expression (2) leads to a free energy

$${1\over N}{\rm log} Z=
\int_{-\pi}^{\pi}\int_{-\pi}^\pi{dk_1\over 2\pi}{dk_2\over 2\pi}{\rm log}
{(\mu^2-4w^2e^{ik_1}{\cos^2{k_2\over 2}}-2\mu e^{ik_1}+e^{2ik_1})}.\en (3)$$

In what follows we first outline the calculation of the density $n_r$ of
lines at site $r$ and of the density-density correlation function. The
probability that a line goes through the site $(r,j)$ is given by

$$n_r=\langle1-\mu\Bpsi_{r,0}\Psi_{r,0}\rangle=n,\en (4)$$

\no where the last equation follows from translational invariance.
The average value in (4) is to be calculated using the partition function given
by
(1) and (3). It is easy to see that $n$ can be expressed in terms of the
Green's
function as

$$n=1+\mu G_{r,r;00}.\en (5)$$

\no The calculation of the Green's function is tedious but straightforward.
Starting from the Fourier components of the Green's function in (2), the
diagonal matrix elements of $G$ read

$$G_{r,r;00}=\int_{-\pi}^{\pi}\int_{-\pi}^\pi{\mu - e^{ik_1}\over(\mu -
e^{ik_1})^2-2w^2e^{ik_1}(1+\cos k_2)}{dk_1\over 2\pi}{dk_2\over 2\pi}.
\en (6)$$

\no By evaluating the double integral in (6) we finally arrive at

$$n=
1-{1\over2\pi}\cos^{-1}\Bigl[1-{(1+\mu)^2\over2(\mu+w^2)}\Bigr]\mp{1\over2\pi}
\cos^{-1}\Bigl[1-{(1-\mu)^2\over2w^2}\Bigr]
\en (7)$$

\noindent Here $\mp$ corresponds, respectively, to cases $\mu>1$ and $\mu<1$.

The result given in (7)
corresponds to the phase diagram shown in Fig.2. $\mu=1\pm 2w$ (A and B
respectively in Fig.2) are lines of phase transitions from a dense system of
directed polymers ($n=1$) to a polymer liquid ($0<n<1$) and finally to a system
without directed polymers. Approaching the phase transition lines from the
regions
II  and III along directions parallel to the $w$-axis, one finds $\partial n/
\partial\mu\sim1/\sqrt{\epsilon}$, where $\epsilon=1-(1-\mu)/2w$ and
$\epsilon=1-(\mu-1)/2w$, respectively for the two cases. There is another
critical phenomenon at $\mu=1$ (broken line in Fig.2). Across this line the
above partial derivative
of the density varies smoothly. This critical behavior can be understood from
the density-density
correlation function $C_{x_2}$, which measure the correlation perpendicular to
the
polymer direction. $C_{x_2}$ is given in terms of the Green's function

$$G_{r,r+xe_2}\equiv G_x=
\int_{-\pi}^{\pi}\int_{-\pi}^{\pi}{(\mu - e^{ik_1})e^{ik_2x}\over(\mu -
e^{ik_1})^2-2w^2e^{ik_1}(1+\cos k_2)}{dk_1\over 2\pi}{dk_2\over 2\pi}
\en (8)$$

\noindent as

$$C_{x}=\mu^2 G_xG_{-x}.\eqno (9)$$

\noindent Performing the $k_1$ integration in (8), one arrives at

$$G_x={1\over 2}\int_{-k^*}^{k^*}{{2\mu (1-\mu)+\alpha}\over
\sqrt{\alpha^2+4\mu\alpha}}e^{ik_2x}{dk_2\over 2\pi}\en (10)$$

\noindent with $k^*=\cos^{-1}[{{(1-\mu)}^2\over{2w^2}}-1]$  and $\a=
2w^2(1+\cos k_2)$. $k^*$ takes the
values $\pi$ and $0$ along the line $\mu=1$ and the transition lines $\mu=1\pm
2w$,
respectively. The calculation of the integral in (10) is an exercise in complex
contour integration, and in the large  $x$ limit can be
performed analytically for any value of $k^*$, with the result

$$G_x\sim\cases{{(-1)^x\over x^2}&for $k^*=\pi$\cr
{\cos({k^*\over 2}){\cos(k^*x)\over x}}&for $k^*\neq \pi$}. \en (11)$$

\noindent Finally, the density-density correlation function decays as $1/x^4$
along the line $\mu=1$ and as $1/x^2$ elsewhere in the liquid phase.

The special property of the line $\mu=1$ is also elucidated by noting that the
expressions for the density of polymer lines, given in (7) are invariant under
a
"duality transformation" $\mu\to 1/\mu$ and $w\to w/\mu$. This duality at
$\mu=1$
reflects the
equivalence of directed polymer lines and lines of empty sites on the lattice.
$\mu=1$ is a self-dual line along which polymers and empty sites compete with
the same
weight. This competition leads to a faster decay of the density-density
correlation
function.

In regions I and IV in Fig. 2, $n$ has the values 1 and 0, respectively. Fig. 3
shows
the behavior of $n$ along the lines $w=0.4$ and $w=0.6$.

If the restrictions imposed on the model as defined above (only steps of length
$2a$ are allowed in the vertical direction, chains must start at $z=0$, at
$r$-sites) are
relaxed, the model describes a system of intersecting random walks. The
analogous
calculations for the partition function (${\tilde Z}$), density of lines
(${\tilde n}$)
and the correlation function (${\tilde C_x}$) are even simpler in this case
(one deals
with a simple Bravais lattice as opposed to a lattice with basis in the case of
non-intersecting walks). We only quote the results

$${1\over N}
{\rm log}{\tilde Z}=\int_{-\pi}^{\pi}\int_{-\pi}^\pi{dk_1\over 2\pi}{dk_2\over
2\pi}{\rm log}{(\mu-2we^{ik_1}\cos k_2-e^{2ik_1})}.\en (12)$$

\noindent Although the above expression for the free energy clearly differs
from (1),
surprisingly,  for the density one obtains  ${\tilde n}\equiv n$, where the
expression for $n$ is
given in (7). This in turn implies that the phase diagram for the IW model is
the sane as the
one obtained for the NIW model.  Finally, the correlation function ${\tilde
C}_x$  can be calculated similarly to $C_x$ in (9). However, instead of $G_x$
in (10) we now have for $\mu=1$
$${\tilde G}_x={1\over 2}\int_{-\pi}^{\pi}{\alpha\over\sqrt{\alpha^2+4}}
e^{ik_2x}{dk_2\over 2\pi}\en (13)$$
with $\alpha=2w\cos k_2$. Considering $\mu=1$ and, for simplicity, $w=1$ and
comparing $G_x$ and ${\tilde G}_x$ we notice that the arguments of the square
root in the denominators are quite different. As a function of $z=e^{ik_2}$
the square root of ${\tilde G}_x$ is analytic in a vicinity of the unit
circle. Therefore, the path of integration can be contracted to a smaller
circle with $|z|<1$. Consequently, ${\tilde G}_x$ decays exponentially on the
$\mu=1$ line. Everywhere else in the liquid phase$^{13}$ it decays like $1/x$.

In conclusion, the analysis of the simple models of non-intersecting and
intersecting directed
polymer chains presented in this work (and studied earlier in references 13 and
14)
leads, surprisingly, to identical thermodynamic properties. The phase diagram
is unusually rich with
multiple
transition lines and phases. There are second order phase transitions, with
diverging second
derivatives of the free energy. The region of the phase diagram between these
transitions, the liquid
phase (A and B in Fig.2) is critical in the sense that in it the
density-density correlation function
decays everywhere according to a power law. An additional feature of the phase
diagram is a self-dual
line in the liquid phase, which manifests itself also in the change of the
decay of correlations. This
change of correlations does not seem to affect the behavior of the
thermodynamic
quantities; they vary smoothly across the self-dual line. The difference
between the two models, in
our study shows up only in the behavior of the correlation function along this
line.

\no One of us (G.F.) acknowledges useful discussions with S. Bhattacharjee.

\vfill
\eject
\noindent{\bf References}
\bigskip
\no
1. V. Privman and N.M. Svrakic, {\it Directed Models of Polymers, Interfaces
and Clusters: Scaling and Finite Size Properties}, Lecture Notes in Physics
338 (Berlin, Springer, 1989)
\bigskip
\no
2. J.J. Lee and G.G. Fuller, J. Coll. Int. Sci. {\bf 103}, 568 (1984)
\bigskip
\no
3. T.E. Harris, J. Appl. Prob. {\bf 2}, 323 (1965); S. Alexander and P. Pincus,
Phys.
Rev. {\bf B18}, 2011 (1978)
\bigskip
\no
4. T.C. Halsey and W. Toor, Phys. Rev. Lett. {\bf 65}, 2820 (1990)
\bigskip
\no
5. M.E. Fisher, J. Stat. Phys. {\bf 34}, 667 (1984)
\bigskip
\no
6. D.R. Nelson, Phys .Rev. Lett.{\bf 60}, 1973 (1988); D.R. Nelson and H.S.
Seung,
Phys.Rev. {\bf B39}, 9153 (1989)
\bigskip
\no
7. K. Ziegler, Europhys.Lett 9, 277 (1989); J. Stat. Phys.{\bf 64}, 277 (1991);
Z.Phys.{\bf B84}, 163 (1991)
\bigskip
\no
8. L.S. Schulman, {\it {Techniques and Applications of Path Integrals}} (John
Wiley \&
Sons, New York, 1981)
\bigskip
\no
9. G. Forgacs, R. Lipowsky and Th.M. Nieuwenhuizen, in {\it{Phase Transitions
and
Critical Phenomena}}, eds. C.Domb and J.L. Lebowitz, Vol. 14 (Academic, London,
1991)
\bigskip
\no
10. J. Nagle, C.S.O. Yokoi and S.M. Bhattacharjee, in {\it { Phase Transitions
and
Critical Phenomena}}, C. Domb and J. Lebowitz eds., Vol.13 (Academic Press,
London 1989)
\bigskip
\no
11. S.M. Bhattacharjee, Europhys.Lett.{\bf 15}, 815 (1991)
\bigskip
\no
12. S.M. Bhattacharjee and J.F. Nagle, Phys. Rev. {\bf A31}, 3199 (1985)
\bigskip
\no
13. E.I. Kornilov and A.A. Litvin, Z.Phys.{\bf B84}, 3 (1991)
\bigskip
\no
14. J.F. Nagle, J.Chem.Phys.63, 1255 (1975)
\bigskip
\no
15. F.A. Berezin, {\it {Method of Second Quantization}} (Academic Press, London
1966)

\vfill
\eject
\noindent{\bf Figure Captions}
\bigskip
\no Fig. 1 The schematic representation of the lattice where the model
presented in this
work is defined. The figure shows a typical configuration of some of the
directed
polymers.The use of periodic boundary conditions in the z direction is
indicated  explicitly.
For more details see text.
\bigskip
\no Fig. 2. The phase diagram of the model. In regions I and IV, $n=1$ and
$n=0$,
respectively.
\bigskip
\no Fig. 3. The behavior of the density of polymer lines along lines $w=0.4$
(full line) and $w=0.6$ (dashed line) in Fig. 2.
\bye